\documentclass[conference]{IEEEtran}

\usepackage[acronym]{glossaries}
\glsdisablehyper

\usepackage[binary-units=true]{siunitx}
\usepackage{subcaption}
\usepackage{algorithm}
\usepackage{xcolor}
\usepackage{listings}
\usepackage{graphicx}
\usepackage{enumerate}
\usepackage{hyperref}
\usepackage{url}
\usepackage{multirow}
\usepackage{booktabs}
\usepackage{csquotes}
\usepackage[inline]{enumitem}
\usepackage{algorithm,algorithmic}
\usepackage{tikz}
\newglossaryentry{A_B^C}{type=main,name={\ensuremath{A_B^C}},sort=A, description={A dummy symbol }}

\newacronym{3gpp}{3GPP}{3rd Generation Partnership Project}
\newacronym{5gppp}{5GPPP}{5G Public Private Partnership}
\newacronym{5qi}{5QI}{5G QoS Identifier}

\newacronym{ack}{ACK}{Acknowledgement}
\newacronym{api}{API}{Application Program Interface}
\newacronym{ap}{AP}{Access Point}
\newacronym{awgn}{AWGN}{Additive White Gaussian Noise}

\newacronym{bssid}{BSSID}{Basic Service Set Identifier}
\newacronym{bss}{BSS}{Basic Service Set}

\newacronym{cca}{CCA}{Clear Channel Assessment}
\newacronym{csma/ca}{CSMA/CA}{Carrier Sensing Multiple Access with Collision Avoidance}
\newacronym{csma}{CSMA}{Carrier Sensing Multiple Access}
\newacronym{cts}{CTS}{Clear to Send}

\newacronym{dcf}{DCF}{Distributed Coordination Function}
\newacronym{difs}{DIFS}{DCF Inter-Frame Spacing}
\newacronym{dl}{DL}{Deep Learning}
\newacronym{dsss}{DSSS}{Direct Sequence Spread Spectrum}

\newacronym{e2e}{E2E}{End-to-End}
\newacronym{edca}{EDCA}{Enhanced Distributed Channel Access}
\newacronym{eirp}{EIRP}{Equivalent Isotropically Radiated Power}
\newacronym{embb}{eMBB}{Enhanced Mobile Broadband}
\newacronym{em}{EM}{Element Management}
\newacronym{epc}{EPC}{Evolved Packet Core}
\newacronym{etsi}{ETSI}{European Telecommunications Standards Institute}

\newacronym{fcs}{FCS}{Frame Check Sequence}

\newacronym{gi}{GI}{Guard Interval}
\newacronym{gui}{GUI}{Graphical User Interface}

\newacronym{ietf}{IETF}{Internet Engineering Task Force}
\newacronym{inp}{InP}{Infrastructure Provider}
\newacronym{iot}{IoT}{Internet of Things}
\newacronym{ism}{ISM}{Industrial Medical Scientific}
\newacronym{itu}{ITU}{International Telecommunication Union}

\newacronym{kpi}{KPI}{Key Performance Indicator}

\newacronym{lan}{LAN}{Local Area Network}

\newacronym{los}{LoS}{Line-of-Sight}

\newacronym{mac}{MAC}{Medium Access Control}
\newacronym{mano}{MANO}{Management and Orchestration}
\newacronym{mcs}{MCS}{Modulation and Coding Scheme}
\newacronym{mimo}{MIMO}{Multiple-Input Multiple-Output}
\newacronym{ml}{ML}{Machine Learning}
\newacronym{mmtc}{mMTC}{Massive Machine-Type Communications}
\newacronym{mpdu}{MPDU}{MAC Protocol Data Unit}
\newacronym{msdu}{MSDU}{MAC Service Data Unit}
\newacronym{mu-mimo}{MU-MIMO}{Multi-User MIMO}

\newacronym{nfvi}{NFVI}{Network Function Virtualization Infrastructure}
\newacronym{nfv}{NFV}{Network Function Virtualization}
\newacronym{nf}{NF}{Network Function}
\newacronym{ngmn}{NGMN}{Next Generation Mobile Networks}
\newacronym{nsi}{NSI}{Network Slice Instance}
\newacronym{ntnu}{NTNU}{Norwegian University of Science and Technology}

\newacronym{ofdma}{OFDMA}{Orthogonal Frequency Division Multiple Access}
\newacronym{ofdm}{OFDM}{Orthogonal Frequency Division Multiplexing}
\newacronym{onf}{ONF}{Open Network Foundation}
\newacronym{os}{OS}{Operating System}

\newacronym{phy}{PHY}{Physical}
\newacronym{psd}{PSD}{Power Spectrum Density}

\newacronym{qoe}{QoE}{Quality of Experience}
\newacronym{qos}{QoS}{Quality of Service}

\newacronym{ran}{RAN}{Radio Access Network}
\newacronym{rl}{RL}{Reinforcement Learning}
\newacronym{rts}{RTS}{Request to Send}
\newacronym{ru}{RU}{Resource Unit}

\newacronym{sdma}{SDMA}{Spatial Division Multiple Access}
\newacronym{sdm}{SDM}{Spatial Division Multiplexing}
\newacronym{sdn}{SDN}{Software Defined Networking}
\newacronym{sdr}{SDR}{Software Defined Radio}
\newacronym{sla}{SLA}{Service Level Agreement}
\newacronym{slo}{SLO}{Service Level Objective}
\newacronym{snr}{SNR}{Signal-to-Noise Ratio}
\newacronym{ssid}{SSID}{Service Set Identifier}
\newacronym{sta}{STA}{Station}

\newacronym{tcp}{TCP}{Transmission Control Protocol}
\newacronym{tpc}{TPC}{Transmit Power Control}

\newacronym{udp}{UDP}{User Datagram Protocol}
\newacronym{ue}{UE}{User Equipment}
\newacronym{urllc}{URLLC}{Ultra-Reliable Low-Latency Communications}

\newacronym{vm}{VM}{Virtual Machine}
\newacronym{vnf}{VNF}{Virtual Network Function}

\newacronym{wdm}{WDM}{Wavelength Division Multiplexing}
\newacronym{wlan}{WLAN}{Wireless Local Area Network}
\newacronym{wsn}{WSN}{Wireless Sensor Network}

\usepackage{comment}

\begin{document}

\title{5G Network Slicing for Wi-Fi Networks}

\author{
\IEEEauthorblockN{Matteo Nerini}
\IEEEauthorblockA{Department of Information Security and\\Communication Technology\\
NTNU -- Norwegian University of Science and Technology\\
Trondheim, Norway\\
Email: matteon@stud.ntnu.no}
\and
\IEEEauthorblockN{David Palma}
\IEEEauthorblockA{Department of Information Security and\\Communication Technology\\
NTNU -- Norwegian University of Science and Technology\\
Trondheim, Norway\\
Email: david.palma@ntnu.no}
}

\maketitle

\begin{tikzpicture}[remember picture,overlay]
\node[anchor=south,yshift=10pt] at (current page.south) {\fbox{\parbox{\dimexpr\textwidth-\fboxsep-\fboxrule\relax}{
\footnotesize \copyright 2021 IEEE.  Personal use of this material is permitted.  Permission from IEEE must be obtained for all other uses, in any current or future media, including reprinting/republishing this material for advertising or promotional purposes, creating new collective works, for resale or redistribution to servers or lists, or reuse of any copyrighted component of this work in other works.\\
\copyright 2021 IFIP. This is the author's version of the work. It is posted here by permission of IFIP for your personal use. Not for redistribution.}}};
\end{tikzpicture}

\begin{abstract}
Future networks will pave the way for a myriad of applications with different requirements and Wi-Fi will play an important role in local area networks.
This is why network slicing is proposed by 5G networks, allowing to offer multiple logical networks tailored to the different user requirements, over a common infrastructure.
However, this is not supported by current Wi-Fi networks.
In this paper, we propose a standard-compliant network slicing approach for the radio access segment of Wi-Fi by defining multiple \glspl{ssid} per \gls{ap}.
We present two algorithms, one that assigns resources according to the requirements of slices in a static way, and another that dynamically configures the slices according to the network's conditions and relevant \glspl{kpi}.
The proposed algorithms were validated through extensive simulations, conducted in the \emph{ns-3} network simulator, and complemented by theoretical assessments.
The obtained results reveal that the two proposed slicing approaches outperform today's Wi-Fi access technique, reaching lower error probability for bandwidth intensive slices and lower latency for time-critical slices.
Simultaneously, the proposed approach is up to 32 times more energy efficient, when considering slices tailored for low-power and low-bandwidth devices, while increasing the overall spectrum efficiency.
\end{abstract}


%
\IEEEpeerreviewmaketitle

\section{Introduction}
\label{sec:introduction}

Network slicing is a technique of paramount importance in the context of 5G.
First introduced in the 5G context by \gls{ngmn}~\cite{ngmn1}, it consists of a virtual and physical division of network resources with customised functionality.
This allows providing logical networks adjusted to the requirements of different use cases on top of a common network.

Network slicing has been widely studied and multiple considerations must be taken into account, such as the used radio access technology and demanded isolation level~\cite{slicing_isolation}.
Multiple radio-access technologies are considered in 5G, including non-\gls{3gpp} technologies such as Wi-Fi, which will play an important role in supporting indoor coverage~\cite{ayy16}.
In fact, \gls{3gpp} considers the interworking with \glspl{wlan} an operational requirement of the next generation access technologies~\cite{3gpp.38.913}.

\gls{3gpp} defines network slicing as the technology that enables the operator to create networks, tailored to provide optimal solutions for different market scenarios which demand diverse requirements, e.g. in terms of functionality, security and \gls{qos}~\cite{3gpp2}.
The \gls{itu}-T provides another definition, stressing the importance of separation between slices, and considers them as logical isolated network partitions composed of multiple virtual resources, isolated and equipped with programmable control and data plane~\cite{itut}.

Network slicing enables allocating different types of resources to serve users according to their needs, simultaneously accommodated over a common infrastructure.
However, implementing network slicing and resource isolation at the radio access presents several challenges and may depend on the used technology.
Using dedicated physical resources (e.g.~radios and antennas) per service allows to implement slicing physical isolation, while other finer-grained physical splitting (e.g. frequency division) allow a better utilisation of resources but may introduce additional challenges (e.g. overhead or interference).

Despite the vast amount of research on slice management and orchestration, network slicing on the \gls{ran} is limited~\cite{ale19}, focusing mostly on \gls{3gpp}'s new radio, with only a few studies on Wi-Fi slicing.
As such, it remains unclear how different 5G classes of service can be supported when resorting to Wi-Fi coverage.
However, the support of multiple \glspl{ssid} is available in current Wi-Fi networks~\cite{zeh17,ale19} and can be used for creating virtual networks on a single \gls{ap}.
This allows the implementation of network slicing for Wi-Fi networks, fully reaching the promised potential of 5G networks even in \gls{wlan} operational environments

The aim of this work is to provide a methodology to efficiently create and maintain network slices in Wi-Fi-based \glspl{ran}, that will accompany 5G in the coming years.
In this paper, we propose a slicing approach compatible with the IEEE 802.11 standard as well as two new algorithms that take into account \glspl{5qi} during slice creation and in real-time for resource management.
Slices are created based on 5G's three main classes of services --- \gls{embb}, \gls{mmtc}, and \gls{urllc}~\cite{itur} --- allocating the necessary resources for each class, according to their requirements (e.g.~error probability, latency, energy efficiency) and the number of served users.

Our results show that the proposed approach outperforms the typical \gls{wlan} single-channel setup.
Grouping wireless stations by classes of services, and dynamically allocating radio resources and changing network parameters of their respective slices, traffic performance is improved as well as the overall resource utilisation.

The proposed solution enables network slicing for current the Wi-Fi radio networks and is extendable to every kind of \gls{csma} communications.
By monitoring important \glspl{kpi} at run-time, the proposed dynamic slicing algorithm is able to save up to 32 times more energy in a realistic \gls{mmtc} scenario and have on average up to one order of magnitude fewer errors in an \gls{embb} scenario. Furthermore, latency for \gls{urllc} devices is decreased by a factor of 5.

The paper is organised as follows. Section~\ref{sec:related_work} covers the background concepts on network slicing applied to the IEEE 802.11 standard.
In Section~\ref{sec:proposal} we describe the Wi-Fi scenario which we want to study, and our proposed slicing solutions.
We assess the performance of our solution in 
Section~\ref{sec:results} followed by concluding remarks in Section~\ref{sec:conclusion}.

\section{Related Work}
\label{sec:related_work}

In today's Wi-Fi networks all the connected \glspl{sta} communicate through only one channel offered by the \gls{ap}.
The IEEE 802.11e amendment introduces the \gls{edca} to assign priority levels to different services.
This channel access technique defines four traffic classes with increasing priority.
However, such classification has limitations since only four traffic classes are available, which do not scale, as identified by Aleixedri et al.~\cite{ale19}.
Furthermore, all the traffic is served by a single \gls{ap}, and there is no guarantee of a certain percentage of airtime for every flow.
In presence of the so-called rate anomaly problem, a traffic flow with high priority could consume too many resources, and optimality could be lost~\cite{tan04}.
Thus, a different methodology is needed to serve prioritised flows.


In order to define \gls{qos}-based slices, we considered these standardised classification methodologies and we adapted them to a 5G network scenario. The different types of 5G services can be grouped in three categories as described next.

\gls{embb} services require support for high data rates and high data traffic volumes.
The main issues taken into account are high user mobility and coverage.
\gls{embb} services are mapped with \gls{5qi} lower than 80~\cite{3gpp3}.
The standardised required throughput for this class of application is \SI[per-mode=symbol]{100}{\mega\bit\per\second} downlink, and \SI[per-mode=symbol]{50}{\mega\bit\per\second} uplink.

Aligned with the vision of the \gls{iot}, \gls{mmtc} services should support a large number of devices that transmit low volumes of delay insensitive data and do not involve mobility.
3GPP does not map this class of services with standardised values of \gls{5qi} due to the low \gls{qos} needed.
However, given the impact that this type of applications will have on future networks, we decided to implement it as a dedicated slice.
The expected density is \SI[per-mode=symbol]{1}{device\per\meter\squared} and the required throughput is below \SI[per-mode=symbol]{100}{\kilo\bit\per\second}~\cite{itur}.
These values are reflected by the most popular \gls{iot} protocols.
For example, LoRa features a raw data rate in the range $10^2$ -- \SI[per-mode=symbol]{e4}{\bit\per\second}~\cite{ade17}, while Narrowband \gls{iot} technology offers $10^2$ -- \SI[per-mode=symbol]{e5}{\bit\per\second}~\cite{rat16}.

\gls{urllc} is the class of services requiring ultra-low latency, high reliability and availability.
\gls{3gpp} defines that \gls{urllc} applications are marked with the highest \glspl{5qi}, from 80 to 86~\cite{3gpp3}.
According to 3GPP's release 15 the required latency for this class of applications is \SI{1}{\milli\second} and the reliability should not be lower than $99.9999\%$.

To date, few studies have been performed on possible network slicing implementations in Wi-Fi contexts.
Most of them are based only on time-scheduled resource allocation, which is proved to be inefficient for resource utilisation.


Aleixendri et al. present a scheduling algorithm to allocate airtime to different users, depending on their \glspl{sla}~\cite{ale19}.
In their approach, each \gls{ap} is equipped with a local scheduler which performs network slicing in time domain.
A global scheduler is also present to manage a group of local schedulers and make adjustments in case some \gls{sla} is violated.
The global scheduler is needed since it is able to manage interference with a broader view of the network.
This algorithm separates resources in orthogonal pools, dividing them in time.
Thus, it is able to ensure a good performance isolation between slices assigning them a minimum of airtime.
The main limitation of this work is that resource utilisation will get worse and network slicing is considered only along the time dimension.
In this way, tailored resources are not assigned to each slice but only a certain percentage of airtime is dedicated.
Furthermore, the work is limited only to downlink transmissions.
While it will be feasible to schedule time resources for downlink connections (since the \gls{ap} has knowledge about the outgoing flows), it will be complex to schedule the uplink traffic in an efficient and practical way, which is not considered.

Richart et al.~\cite{ric17} present a Wi-Fi slicing solution where the objective is to allocate the airtime resource to different slices.
In their approach, when a slice requests a percentage of the total airtime to an \gls{ap}, if enough resources are available, the airtime is allocated to the slice.
A similar methodology based on airtime slicing was studied by Zehl et al.~\cite{zeh17}.
The presented methodology provides complete traffic separation in downlink between the home network and the so-called guest network.
However, traffic is scheduled only for the downlink and the time division of resources does not provide advantages from the resource utilisation point of view.

De Bast et al. present a \gls{dl} algorithm to enable network slicing considering a radio access segment of an IEEE 802.11ac network and aim at assigning the proper resources for each user depending on its required throughput~\cite{deb19}.
The resources managed by the algorithm are: the \gls{mcs} index, the \gls{cca} threshold, and the transmission power.
The main limitation of this approach is that all the \glspl{sta} are connected to the same channel and will interfere with each other.
In this way, a good performance isolation will never be reached.
Furthermore, the used \gls{dl} technique requires an extensive pre-training and a long time to converge on the optimal solution.
This is not likely in a realistic scenario in which scheduling decisions must happen at run-time.

Makhlouf et al. propose a totally new access scheme to realise network slicing \cite{mak14}.
They observe that the main reason for the inefficiency of Wi-Fi standards is that the \gls{mac} Layer randomly allocates the whole channel available to only one user as a single resource.
For this reason they divide the entire channel in different sub-channels, assigning them to the users according to their requirements and conditions.
However, this approach is not standard compliant.

Finally, Gu et al. propose to realise a multi-tenant architecture on a single \gls{ap} by installing different \glspl{ssid} on it~\cite{gu18}.
The objective is to create coexisting separated networks and increase the overall network throughput.
Network slicing is not realised because each tenant is served with an opportunistic approach and users are not differentiated by their requirements.
However, exploiting multiple \glspl{ssid} on a single \gls{ap} opens up for new possibilities. We want to apply this idea to overcome the above mentioned limitations of previous works.

\section{Wi-Fi SSID Network Slicing}
\label{sec:proposal}

In this paper we focus on slicing at the radio access segment of the network and consider a Wi-Fi \gls{ap} providing different services to associated \glspl{sta} in an indoor environment.
We propose the allocation of different radio channels, with different characteristics (e.g.~bandwidth and centre frequency), each dedicated to a pre-defined slice.
The channels are identified with distinct \gls{ssid}, and treated as separated networks.
Thus, the \glspl{sta} associated with the \gls{ap} access the relevant channel through a \gls{csma/ca} scheme, as in IEEE 802.11.

We take into consideration the uplink communication between the \glspl{sta} and the \gls{ap}, tackling the difficulties created by distributed resource sharing.
This differentiates our work from related studies, which focus only on the downlink scheduling of resources.
In a downlink scenario, the \gls{ap} has full control and it can serve multiple users with well known broadcast techniques (e.g.~space-time coding and beamforming if a multi-antenna system is available)~\cite{lei02}.
On the other hand, in uplink transmission the inputs are located in independent devices which do not have a broad view of the network.

In our approach we propose the definition of three slices, based on 5G's three main scenarios: Slice A for \gls{embb}, Slice B for \gls{mmtc} and Slice C for \gls{urllc}.
This is achieved by defining the three channels, identified by their \glspl{ssid}, and assigning them a bandwidth, a centre frequency (given by the channel number), and deciding the \gls{gi}, the \gls{mcs} index and the transmission power $P_{TX}$.
By assigning different \glspl{sta} to different channels an improved control of shared resources is achieved, including in the uplink.

The traditional access methodology implemented in today's Wi-Fi networks connects all \glspl{sta} to a single channel.
In an optimistic setting a channel bandwidth of \SI{160}{\mega\hertz} could be considered to support the maximum throughput.
Similarly, the highest transmission power allowed in Europe is \SI{20}{dBm} \cite{etsi5}.
This allows us to obtain the best \gls{snr} but can create interference problems when considering multiple \glspl{sta}.
A high \gls{gi} can be used to decrease inter-symbol interference but it increases latency.
Another important consideration is the chosen \gls{mcs}, which will provide a trade-off between coverage and maximum achievable throughput, since higher \glspl{mcs} will not allow distant devices to properly communicate due to an insufficient \gls{snr}~\cite{ner20}.

\subsection{Static Slicing Algorithm}

\begin{algorithm}[t!]
\caption{Static Algorithm Slice A (eMBB)}
\label{alg:static_A}
\begin{algorithmic}[1]
\small
\algsetup{linenosize=\scriptsize}
\renewcommand{\algorithmicensure}{\textbf{Output:}}
\ENSURE $chWidthA, chNumberA, giA, mcsA, txPowerA$
\STATE $chWidthA = CB_{wmin}(dataRateA, mcsA, giA)$;
\STATE $chNumberA = chNumberA(chWidthA)$;
\STATE $giA =$ \SI{1600}{\nano\second}; $mcsA = 5$; $txPowerA =$ \SI{20}{dBm};
\end{algorithmic} 
\end{algorithm}

\begin{algorithm}[t!]
\caption{Static Algorithm Slice B (mMTC)}
\label{alg:static_B}
\begin{algorithmic}[1]
\small
\algsetup{linenosize=\scriptsize}
\renewcommand{\algorithmicensure}{\textbf{Output:}}
\ENSURE $chWidthB, chNumberB, giB, mcsB, txPowerB$
\STATE $chWidthB =$ \SI{20}{\mega\hertz}; $chNumberB = 100$;
\STATE $giB =$ \SI{1600}{\nano\second}; $mcsB = 5$; $txPowerB =$ \SI{20}{dBm};
\end{algorithmic} 
\end{algorithm}

\begin{algorithm}[t!]
\caption{Static Algorithm Slice C (URLLC)}
\label{alg:static_C}
\begin{algorithmic}[1]
\small
\algsetup{linenosize=\scriptsize}
\renewcommand{\algorithmicensure}{\textbf{Output:}}
\ENSURE $chWidthC, chNumberC, giC, mcsC, txPowerC$
\STATE $chWidthC = CB_{wmin}(dataRateC, mcsC, giC)$;
\STATE $chNumberC = chNumberC(chWidthC)$;
\STATE $giC =$ \SI{1600}{\nano\second}; $mcsC = 5$; $txPowerC =$ \SI{20}{dBm};
\end{algorithmic} 
\end{algorithm}

Our static network slicing algorithm creates three separate channels, one for each of the defined slice.
These channels are placed in the \SI{5}{\giga\hertz} region in order to have the maximum free distance between each other.
More precisely, the channel number for Slice A is as low as possible depending on the bandwidth (function $chNumberA(chWidthA)$ in the algorithms); Slice B is allocated in the centre of the \SI{5}{\giga\hertz}, with channel number 100; and the channel for Slice C has the highest channel number, thus minimising inter-channel interference: function $chNumberC(chWidthC)$.

The channel bandwidth is defined considering the sum of the required throughput in each slice for a given number of \glspl{sta}.
The \glspl{gi}, the \gls{mcs} indexes and the transmission powers have the same fixed values for all slices (\SI{1600}{\nano\second}, 5 and \SI{20}{dBm}).
These are estimates that intend to cover most cases.




Algorithms~\ref{alg:static_A}, \ref{alg:static_B} and \ref{alg:static_C} show how resources are allocated.
The variable $dataRateX$ represents the total throughput required in the slice $X$, where $X \in \{A, B, C\}$, and the function $CB_{wmin}(dataRateX,MCS,GI) = min\{CB_w \in \{20, 40, 80, 160\} \mid dataRate(CB_w, MCS, GI) \ge dataRateX\}$ corresponds to the required channel bandwidth.
For example, assuming that an \gls{mmtc} device in Slice B can require at most \SI[per-mode=symbol]{50}{\kilo\bit\per\second}, the bandwidth is fixed to \SI{20}{\mega\hertz} (the lowest possible in IEEE 802.11), which can achieve \SI[per-mode=symbol]{58}{\mega\bit\per\second} with $MCS=5$ and $GI=\SI{1600}{\nano\second}$.

\subsection{Dynamic Slicing Algorithm}
The dynamic slicing algorithm starts by assigning a channel per slice, which is then continuously monitored and updated at run-time, at a time interval $T$.
The algorithm exploits the network conditions and considers the reached performance to adapt slices to the network's needs.
Three algorithms have been created for each slice type because of their distinct needs.

%
%
%
%

\subsubsection{Dynamic Algorithm for Slice A (eMBB)}


\begin{algorithm}[t]
\caption{Dynamic Algorithm Slice A (eMBB)}
\label{alg:dynamic_A}
\begin{algorithmic}[1]
\small
\algsetup{linenosize=\scriptsize}
\renewcommand{\algorithmicensure}{\textbf{Output:}}
\ENSURE $chWidthA, chNumberA, giA, mcsA, txPowerA$
\\ \textit{Initialisation}:
\STATE $txPowerA = \SI{20}{dBm}$; $giA = \SI{800}{\nano\second}$;
\STATE $mcsA_{max} = MCS_{max}(rxPowerA_{min})$;
\STATE $chWidthA = CB_{wmin}(dataRateA, mcsA_{max}, giA)$;
\STATE $chNumberA = chNumberA(chWidthA)$;
\STATE $mcsA_{min} = MCS_{min}(dataRateA, chWidthA, giA)$;
\STATE \textbf{if} $mcsA_{max} > mcsA_{min}$ \textbf{then} $mcsA = mcsA_{min} + 1$;
\STATE \textbf{else} $mcsA = mcsA_{max}$;
\\ \textit{LOOP Process}:
\STATE \textbf{if} SLA KO in last $T$ \&\& $P_e$ in last $T$ $>$ in prev. $T$ \textbf{then} $chWidthMulA = 2$;
\STATE \textbf{if} SLA OK in last $T$ \&\& $P_e$ in last $T$ $<$ in prev. $T$ \textbf{then} $chWidthMulA = 1$;
\STATE $mcsA_{max} = MCS_{max}(rxPowerA_{min})$;
\STATE $chWidthA = CB_{wmin}(dataRateA, mcsA_{max}, giA)$;
\STATE $chWidthA = chWidthA \times chWidthMulA$;
\STATE $chNumberA = chNumberA(chWidthA)$;
\STATE $mcsA_{min} = MCS_{min}(dataRateA, chWidthA, giA)$;
\STATE \textbf{if} $mcsA_{max} > mcsA_{min}$ \textbf{then} $mcsA = mcsA_{min} + 1$;
\STATE \textbf{else} $mcsA = mcsA_{max}$;
\end{algorithmic} 
\end{algorithm}


Slice A is initialised according to the first part of Algorithm~\ref{alg:dynamic_A}.
$P_{TX}$ is set to \SI{20}{dBm} in order to use higher \gls{mcs} indexes and to enable higher data rates.
For the same reason, the \gls{gi} is set to the standard's lowest possible value $giA = \SI{800}{ns}$.

Before starting, the \gls{ap} obtains the received power for every connected device.
Thus, it finds the minimum value, $rxPowerA_{min}$, corresponding to one of the farthest \glspl{sta}.
Then, the maximum achievable \gls{mcs}, $mcsA_{max}$, is computed with the function $MCS_{max}(P_{RX}) = max\{MSC \in \{0,\dots , 11\} \mid P_e(MCS, P_{RX}) < 0.001\}$.
This allows determining the maximum allowed \gls{mcs} for a given received power $P_{RX}$ for a packet error probability ($P_e$) lower than $0.001$.
Now, the function $CB_{wmin}(dataRateA, mcsA_{max}, giA)$ is called to obtain the minimum viable channel bandwidth, $chWidthA$.
Finally, having fixed the channel bandwidth, the minimum \gls{mcs} that still supports the required throughput is determined with the function $MCS_{min}(dataRateA, chWidthA, giA)$.

%
%
%
%

After the initialisation phase, the slice is updated according to the network conditions every interval $T$.
The packet error probability experienced in each flow during the last interval $T$ is calculated and compared with a threshold:
\begin{equation}
P_{e} = \frac{txPackets - rxPackets}{txPackets} \stackrel{?}{\leq} 0.02.
\label{eqn:slaA}
\end{equation}

\begin{figure}
\centering
\includegraphics[width=0.9\columnwidth]{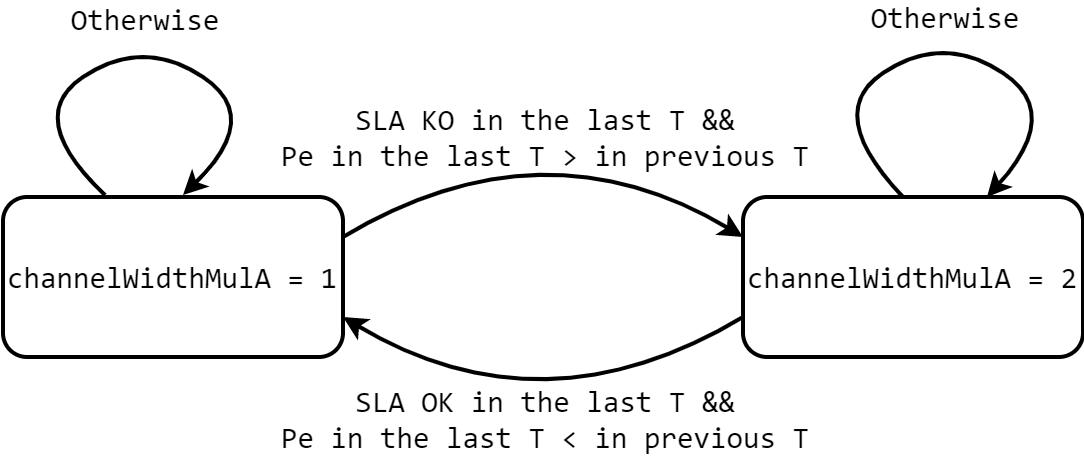}
\caption{State machine implemented by the dynamic algorithm for resource allocation in Slice A (eMBB).}
\label{fig:alg_sliceA}
\end{figure}

The threshold was set to $0.02$ after preliminary experiments based on a typical single channel approach, which showed that only a tenth of the \glspl{sta} reached an error probability lower than 0.02.
We used this value as our \gls{sla} and based on Equation~\ref{eqn:slaA} the algorithm can make a decision about the required bandwidth for the next interval of time.
The defined behaviour is presented by the state machine in Figure~\ref{fig:alg_sliceA}.

Every $T$, the needed bandwidth is recomputed and multiplied by the value $chWidthMulA$ which can be $1$ or $2$ whether or not the bandwidth needs to be duplicated.
A transition from state $1$ to state $2$ occurs if Equation~\ref{eqn:slaA} does not hold for at least one \gls{sta}, and the total error probability in the last $T$ is worse.
A transition from state $2$ to state $1$ occurs if Equation~\ref{eqn:slaA} holds for every \gls{sta} and the total error probability in the last $T$ is better.
Once the bandwidth has been fixed, the \gls{mcs} is determined.

\subsubsection{Dynamic Algorithm for Slice B (mMTC)}
Slice B is initialised as seen in Algorithm~\ref{alg:dynamic_B}.
The objective of this algorithm is to minimise the transmission power since energy saving is crucial in \gls{iot} applications and \gls{mmtc} communications.

\gls{mmtc} has low throughput requirements, thus the channel bandwidth for this slice was fixed to \SI{20}{\mega\hertz}, the smallest possible in IEEE 802.11.
The channel number is 100, at the centre of the \SI{5}{\giga\hertz} region far from the other two slices.
Also the \gls{gi} is fixed and can be relaxed to \SI{1600}{\nano\second} since there are no strict requirements on latency.
The lowest \gls{mcs} is selected according to the function $MCS_{min}(dataRate,CB_w,GI)$, with an additive margin $mcsAddB$ initialised to 1.
Once the \gls{mcs} is fixed, the minimum received power is found with
$P_{RXmin}(MCS) = min\{P_{RX} \mid P_e(MCS, P_{RX}) < 0.001\}$,
the inverse of $MCS_{max}(P_{RX})$.
From this value and the path losses experienced by the users, the minimum usable transmission power is found\footnote{Path loss is obtained by the \gls{ap} by fixing \glspl{sta}' signalling messages transmission power. By measuring the received power, we will have: $loss = P_{TX} - P_{RX}$ $[dB]$.}.

In Algorithm~\ref{alg:dynamic_B}, the transmission power is set to accommodate the requirements of 90\% of the users.
We consider that having re-transmissions for a tenth of the users in this slice is not an issue.
The data volumes are small and with stringent latency needs.
Therefore, by increasingly sorting the registered losses ($lossB$ in the algorithm), it is possible to calculate the required $P_{TX}$ for Slice B.
An additive margin $txPowerAddB$ (initialised to 3) is added to the minimum power needed.
As explained below, this parameter is modified at run-time to adjust the transmission power based on the obtained \gls{qos}.

%
%

%
%


\begin{algorithm}[t]
\caption{Dynamic Algorithm Slice B (mMTC).}
\label{alg:dynamic_B}
\begin{algorithmic}[1]
\small
\algsetup{linenosize=\scriptsize}
\renewcommand{\algorithmicensure}{\textbf{Output:}}
\ENSURE $chWidthB, chNumberB, giB, mcsB, txPowerB$
\\ \textit{Initialisation}:
\STATE $mcsAddB = 1$; $txPowerAddB = 3$;  $giB = \SI{1600}{\nano\second}$;
\STATE $chWidthB = \SI{20}{\mega\hertz}$; $chNumberB = 100$;
\STATE $mcsB = MCS_{min}(dataRateB, chWidthB, giB)$;
\STATE $mcsB = mcsB + mcsAddB$;
\STATE $txPowerB = lossB[9nStaB/10] + P_{RXmin}(mcsB)$;
\STATE $txPowerB = txPowerB + txPowerAddB$;
\\ \textit{LOOP Process}:
\IF {SLA KO in last $T$ \&\& $P_e$ in last $T$ $>$ in prev. $T$}
\STATE \textbf{if} $txPowerAddB < 6$ \textbf{then} $txPowerAddB++;$
\STATE \textbf{else if} $mcsAddB < 4$ \textbf{then}
\\ $mcsAddB++$; $txPowerAddB = 3$;
\ENDIF
\IF {SLA OK in last $T$ \&\& $P_e$ in last $T$ $<$ in prev. $T$} 
\STATE \textbf{if} $txPowerAddB > 1$ \textbf{then} $txPowerAddB--;$
\STATE \textbf{else if} $mcsAddB > 1$ \textbf{then}
\\ $mcsAddB--$; $txPowerAddB = 3$;
\ENDIF
\STATE $mcsB = MCS_{min}(dataRateB, chWidthB, giB)$;
\STATE $mcsB = mcsB + mcsAddB$;
\STATE $txPowerB = lossB[9nStaB/10] + P_{RXmin}(mcsB)$;
\STATE $txPowerB = txPowerB + txPowerAddB$;
\end{algorithmic} 
\end{algorithm}


\gls{mmtc} slices are updated according to the network conditions on a regular interval $T$, by calling the \textit{LOOP Process} in Algorithm~\ref{alg:dynamic_B}.
The error probability is calculated for each flow as in Equation~\ref{eqn:slaA} and compared with the threshold 0.2.
If Equation~\ref{eqn:slaA} does not hold in at least 90\% of the \glspl{sta}, and the average error probability is getting worse with respect to the previous interval $T$, the algorithm reacts by increasing the margin $txPowerAddB$.
When it reaches its maximum value, the $mcsAddB$ margin is also increased.
In this way, the time on air required for each transmission will decrease and the number of collisions due to interference will be reduced, requiring however a higher transmission power.
When Equation~\ref{eqn:slaA} is satisfied in 90\% of the \glspl{sta}, and the overall error probability is lower than in the previous interval $T$, the transmission power can be decreased.
In a dual way, the algorithm acts on the parameters $txPowerAddB$ and $mcsAddB$ to relax the assigned resources.

%
%
%
%

\subsubsection{Dynamic Algorithm for Slice C (URLLC)}

\begin{algorithm}[t]
\caption{Dynamic Algorithm Slice C (URLLC).}
\label{alg:dynamic_C}
\begin{algorithmic}[1]
\small
\algsetup{linenosize=\scriptsize}
\renewcommand{\algorithmicensure}{\textbf{Output:}}
\ENSURE $chWidthC, chNumberC, giC, mcsC, txPowerC$
\\ \textit{LOOP Process}:
\STATE $txPowerC = \SI{20}{dBm}$; $giC = \SI{800}{\nano\second}$;
\STATE $mcsC_{max} = MCS_{max}(rxPowerC_{min})$;
\STATE $chWidthC = CB_{wmin}(dataRateC, mcsC_{max}, giC)$;
\STATE $chNumberC = chNumberC(chWidthC)$;
\STATE $mcsC_{min} = MCS_{min}(dataRateC, chWidthC, giC)$;
\STATE \textbf{if} $mcsC_{max} > mcsC_{min}$ \textbf{then} $mcsC = mcsC_{max}$;
\STATE \textbf{else} $mcsC = mcsC_{min} + 1$;
\STATE $chNumberC = chNumberC(chWidthC)$;
\end{algorithmic} 
\end{algorithm}


Slice C is initialised as reported in the \emph{LOOP Process} of Algorithm~\ref{alg:dynamic_C}.
It is similar to the initialisation for Slice A, except for the selection of channel numbers and the \gls{mcs} selection method.
The channel numbers in Slice C are selected to be the highest possible in order to allocate channels as distant as possible from each other.
In this way, Slice A and Slice C are situated at the extremes of the \SI{5}{\giga\hertz} band, and Slice B exactly in the middle.
Finally, the desired \gls{mcs} is also obtained in a different way.
In Slice A, the minimum \gls{mcs} able to support the required throughput is chosen, here we select the highest one usable according to the received powers.
This difference is due to the fact that a higher \gls{mcs} should decrease latency, an important aspect of the \gls{urllc} slice.
However, this may lead to increased error probability, which will have to be carefully handled.

At every interval of time $T$, Algorithm~\ref{alg:dynamic_C} is called to update the \gls{urllc} slice.
The goal is to always recompute the highest performing \gls{mcs} with the maximum possible transmission power.
Experimentally, we found that doubling the channel bandwidth at real time, in order to fulfil performance requirements, does not bring any positive effects and worsens resource utilisation.
As such, adjusting the \gls{mcs} to the highest cardinality should be the focus, allowing for lower latencies.

%
%
%

\begin{figure*}[t]
\centering
    \begin{subfigure}[b]{0.32\textwidth}
    \centering
    \includegraphics[width=\columnwidth]{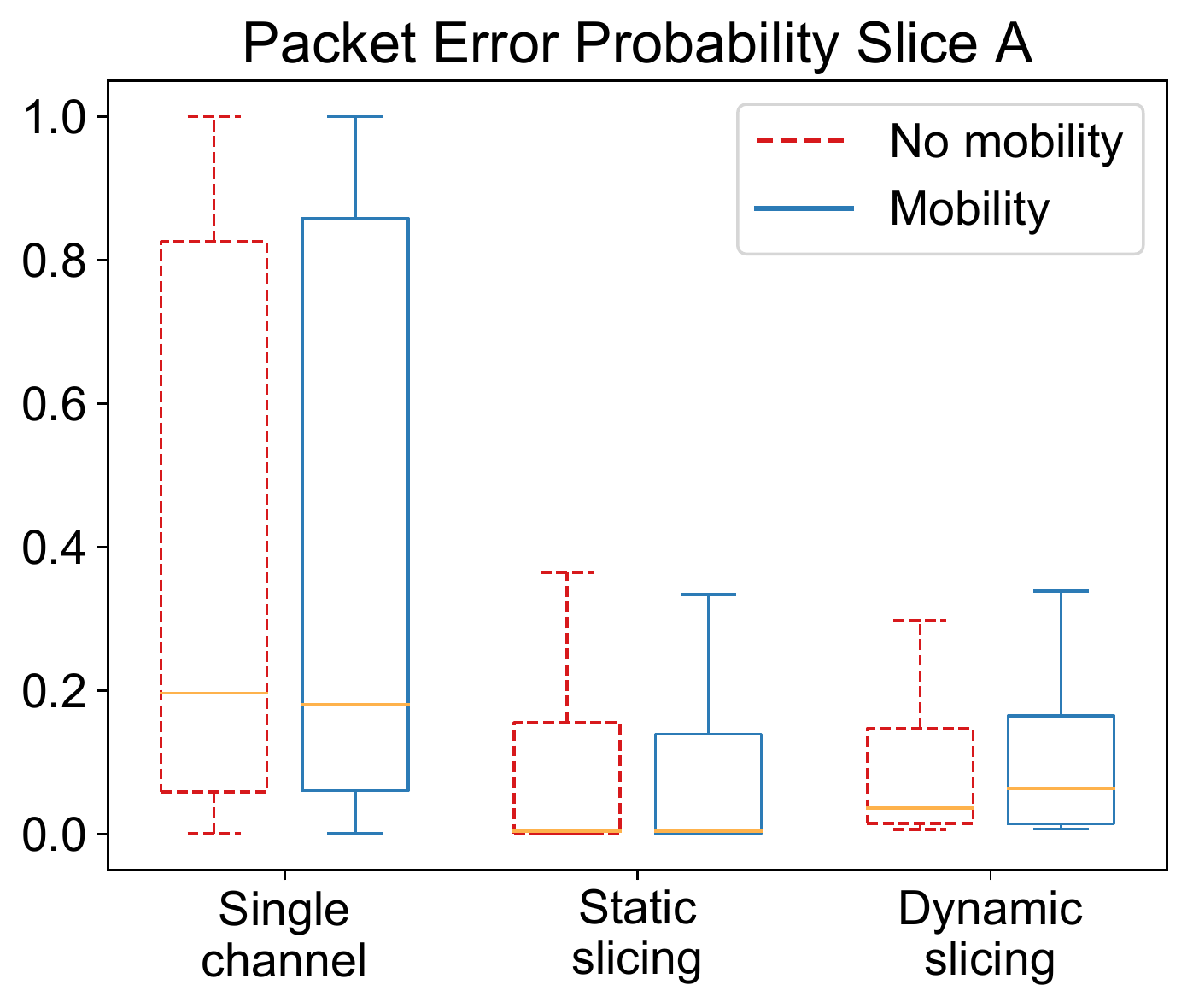}
    \caption{eMBB}\label{fig:peA}
    \end{subfigure}
    \begin{subfigure}[b]{0.327\textwidth}  
    \centering 
    \includegraphics[width=\columnwidth]{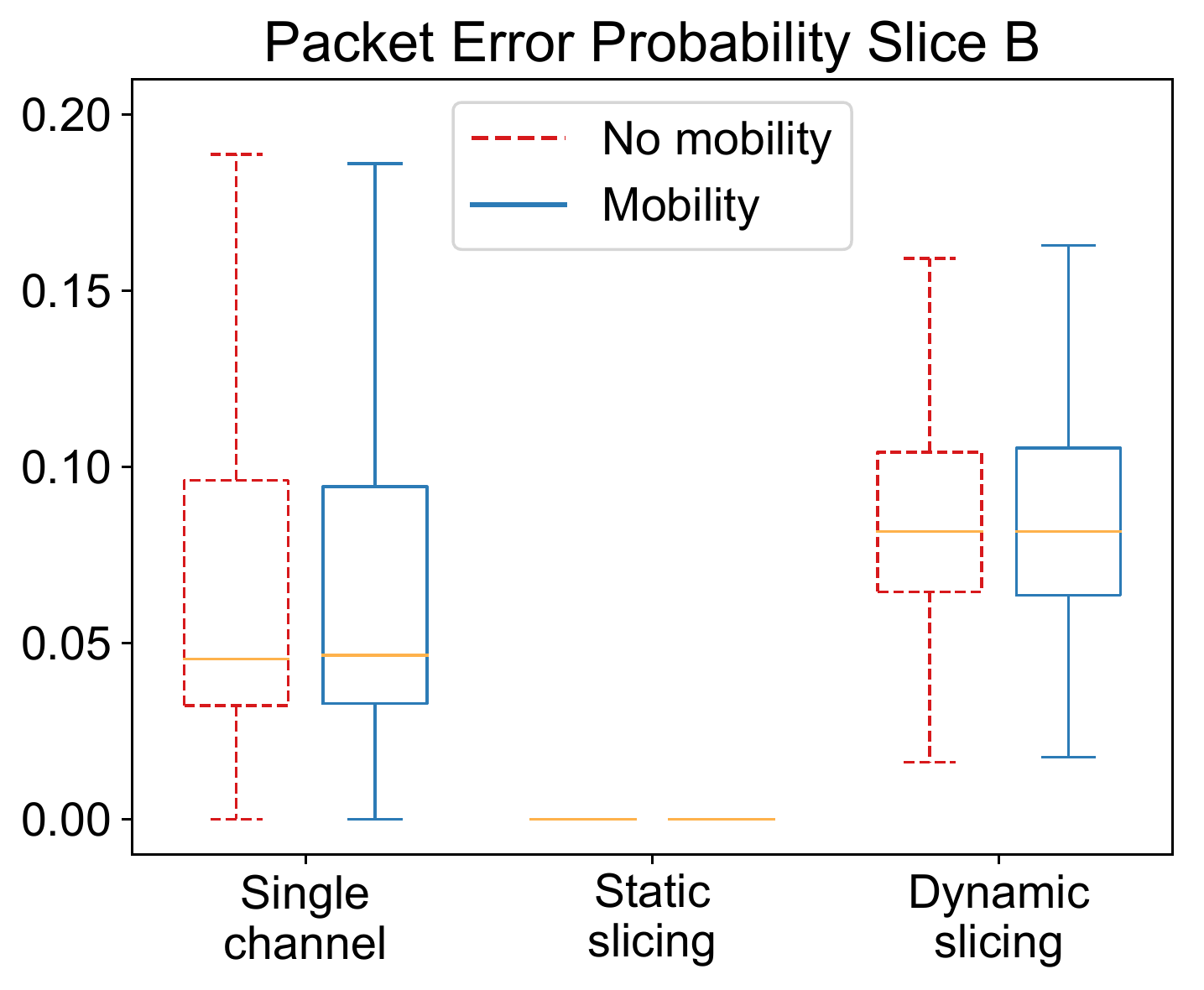}
    \caption{mMTC}\label{fig:peB}
    \end{subfigure}
    \begin{subfigure}[b]{0.32\textwidth}   
    \centering 
    \includegraphics[width=\columnwidth]{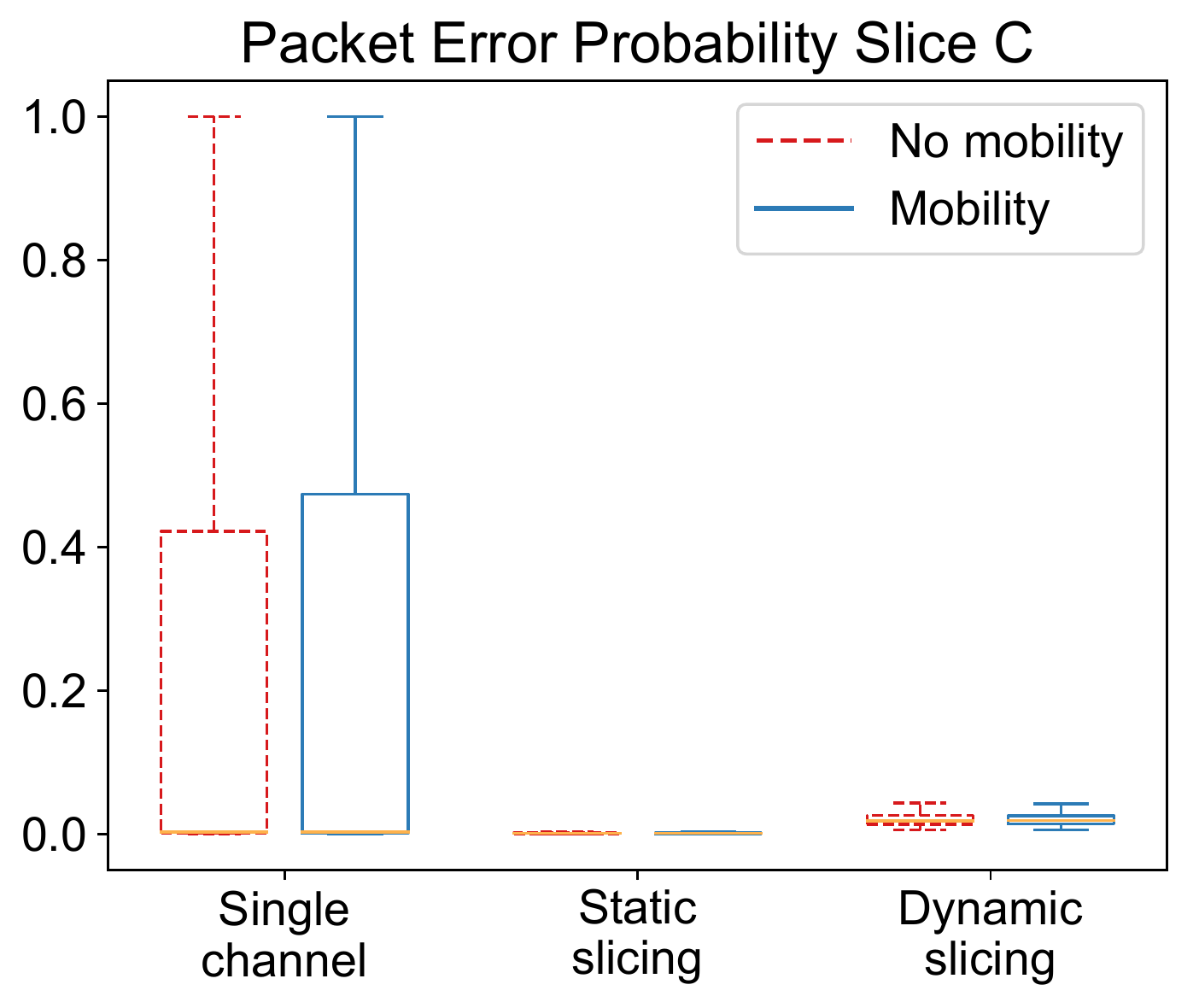}
    \caption{URLLC}\label{fig:peC}
    \end{subfigure}
\caption{Packet error probability for the three slices A, B and C.
}
\label{fig:pe}
\end{figure*}

\subsection{ns-3 Simulation Setup and Methodology}

In this study, we used \emph{ns-3} simulator due to its CPU utilisation, memory usage, computational time, and scalability as a wireless network simulator~\cite{kha14}.
We utilised the latest Wi-Fi version supported by \emph{ns-3} (IEEE 802.11ax), which does not yet support the \gls{ofdma} and that could be used to explore different slicing techniques.
Therefore, we focused on \gls{csma/ca} access.
In addition, directional antennas are not yet supported and therefore \gls{mu-mimo} and beamforming were not considered.
\emph{ns-3} has also been used for its capability to simulate frequency dependent effects in a realistic way (e.g.~the cross-channel interference).
In the default physical layer model, \emph{YansWifiPhy}~\cite{lac06}, two Wi-Fi channels interfere with each other only if they have the same channel number, and interference from non-Wi-Fi technologies is not considered.
However, to guarantee an accurate assessment we employed the \emph{SpectrumWifiPhy} model, which provides a realistic representation of radio links.
\begin{table}
\centering
\caption{Parameters setting for all the simulations.}
\begin{tabular}{@{}ll@{}}
\toprule
\textbf{Parameter} & \textbf{Value}\\
\midrule
Data Rate eMBB      & $\sim U[80,100]$ Mbit/s\\
Data Rate mMTC      & $\sim U[30,50]$ Kbit/s\\
Data Rate URLLC     & $\sim U[20,40]$ Mbit/s\\
Data Retransmission & none\\
Packet Size         & \SI{1472}{Bytes}\\
\emph{ns-3} PHY Layer Model     & \emph{SpectrumWifiPhy}\\
Positions $(x,y)$   & $X \sim U[0, 20]$ m, $Y \sim U[0, 10]$ m\\
Simulation Time     & \SI{15}{\second}\\
Standard -- Band    & 802.11ax -- \SI{5}{\giga\hertz}\\
\bottomrule
\end{tabular}
\label{tab:wifin}\vspace{-10pt}
\end{table}


We consider an indoor Wi-Fi scenario consisting in multiple \glspl{sta} associated to one \gls{ap}.
All the \gls{sta} devices are randomly placed in a rectangular room with dimensions 20 $\times$ \SI{10}{\meter}, \SI{1.5}{\meter} above the ground.
The \gls{ap} is situated at the centre of the room at the ceiling level, \SI{3}{\meter} high.

Our scenario includes between 2 and 6 \gls{embb} STAs.
Mobility is taken into account by moving the \glspl{sta} of this category at a pedestrian speed with variable direction.
This represents a typical mobility pattern of an indoor user connected to Wi-Fi~\cite{ayy16}.
For each \gls{embb} device, the required uplink throughput is random, uniformly distributed between 80 and \SI[per-mode=symbol]{100}{\mega\bit\per\second}.
We consider 100 \gls{mmtc} devices, in a fixed randomised position, connected to the \gls{ap} to reflect a density of about \SI[per-mode=symbol]{1}{device\per\meter\squared}.
Each of them transmits data at a random speed in the range \SIrange[per-mode=symbol,range-units=single,range-phrase=--]{30}{50}{\kilo\bit\per\second}.
Finally, our simulations include 2 to 6 \gls{urllc} devices with a random throughput spanning from \SIrange[per-mode=symbol,range-units=single]{20}{40}{\mega\bit\per\second}, which also follow an indoor mobility pattern.


In the \emph{ns-3} environment, it is possible to simulate the presence of buildings and rooms with the \emph{Buildings Module}.
Using this model we created a single room and located devices inside it.
The \emph{HybridBuildingsPropagationLossModel} has been used as propagation loss model and \emph{RandomWalk2dMobilityModel} was used to add a pedestrian movement to nodes in slices A and C.
In our simulations, the direction of each mobile device is uniformly chosen every second within the interval [0,$2\pi$] radians and the speed in the range \SIrange[range-phrase=--,per-mode=symbol,range-units=single]{2}{4}{\meter\per\second}.
If they hit one of the boundaries (specified by the 20 $\times$ \SI{10}{\meter} rectangle), they rebound with a reflexive angle and speed.
Each simulation setting was run 20 times with a different seed.
Finally, in our simulations we set $T = \SI{1}{\second}$, for invoking the dynamic algorithms.
For the sake of reproducibility, Table~\ref{tab:wifin} summarises the used parameters for every carried out experiment.

In order to consider different proportions of users in the defined slices, and to study the robustness of the proposed algorithms, three simulation settings were considered with a different number of \glspl{sta} per slice:
\begin{itemize}
\item \textbf{2-100-6}: $nStaA = 2$, $nStaB = 100$, $nStaC = 6$;
\item \textbf{4-100-4}: $nStaA = 4$, $nStaB = 100$, $nStaC = 4$;
\item \textbf{6-100-2}: $nStaA = 6$, $nStaB = 100$, $nStaC = 2$.
\end{itemize}
where $nStaA$, $nStaB$ and $nStaC$ denote the number of devices in Slice A (\gls{embb}), B (\gls{mmtc}) and C (\gls{urllc}).

The required throughput and the initial position of each user was randomised at every run and the simulation duration was \SI{15}{\second}, since a higher simulation time did not influence the performance of the 108 concurrent data flows.
Furthermore, even at the lowest speed setting (\SI[per-mode=symbol]{2}{\meter\per\second}), nodes could move across the entire room in \SI{15}{\second}, which should be representative of different radio conditions.\footnote{All of the source code, including the \emph{ns-3} scripts involved, is available at
\href{https://github.com/matteonerini/5g-network-slicing-for-wifi-networks}
{https://github.com/matteonerini/5g-network-slicing-for-wifi-networks}.}

\section{Results}
\label{sec:results}

In this section, we present the results obtained by simulating the three resource allocation algorithms.
The first one, the \emph{Single channel approach} represents typical Wi-Fi networks exploiting a channel width of \SI{160}{\mega\hertz} on channel number 50, \gls{mcs} 5, a \gls{gi} of \SI{1600}{\nano\second} and a transmission power of TX power of \SI{20}{dBm}.
The second, \emph{Static network slicing}, allocates resources at the beginning of the simulation without any change.
The third, \emph{Dynamic network slicing}, schedules resources at every time interval $T$, to adjust slices based on the achieved performance and the network conditions.

\begin{figure*}[t]
\centering
    \begin{subfigure}[b]{0.327\textwidth}
    \centering
    \includegraphics[width=\columnwidth]{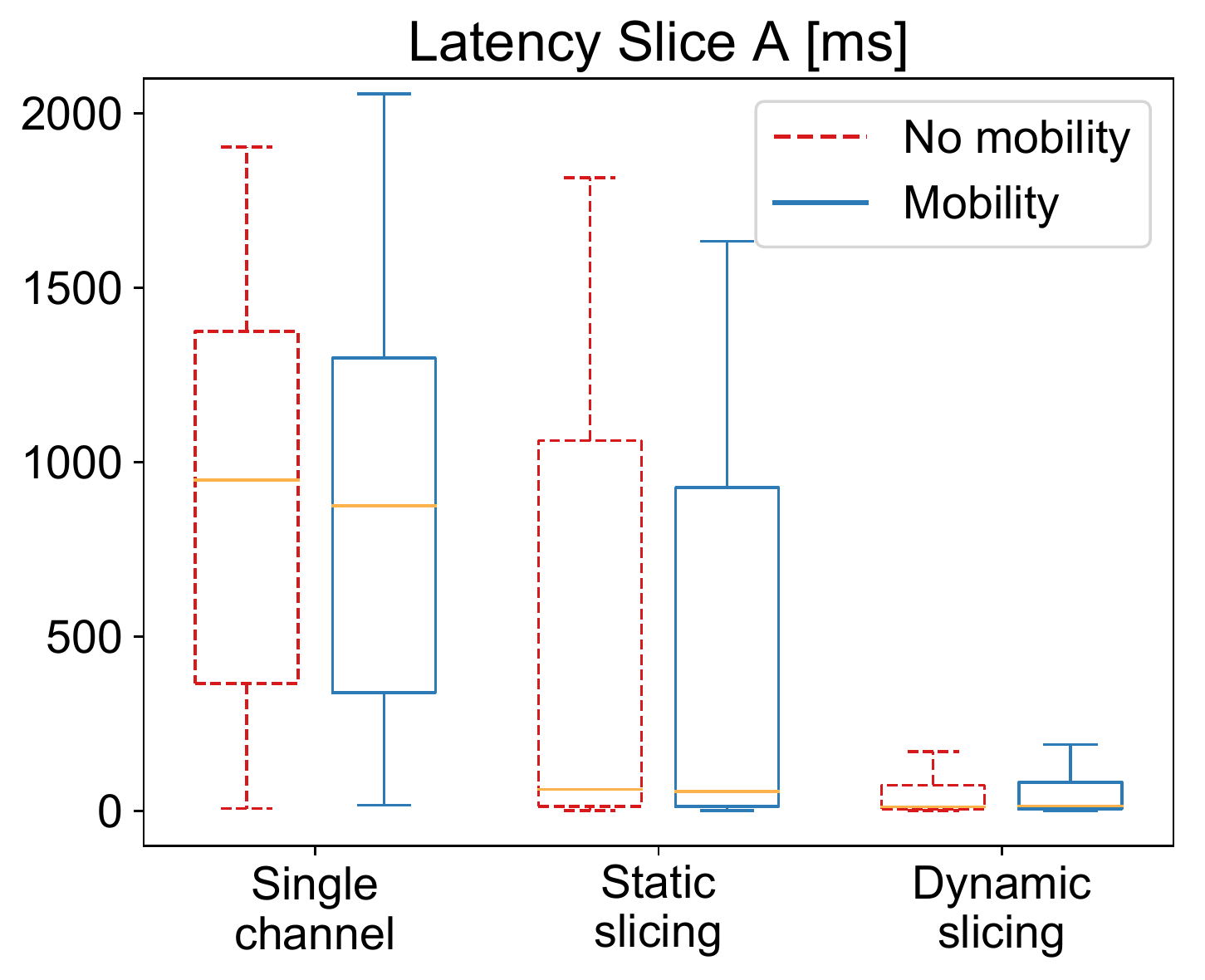}
    \caption{eMBB}\label{fig:latA}
    \end{subfigure}
    \begin{subfigure}[b]{0.32\textwidth}  
    \centering 
    \includegraphics[width=\columnwidth]{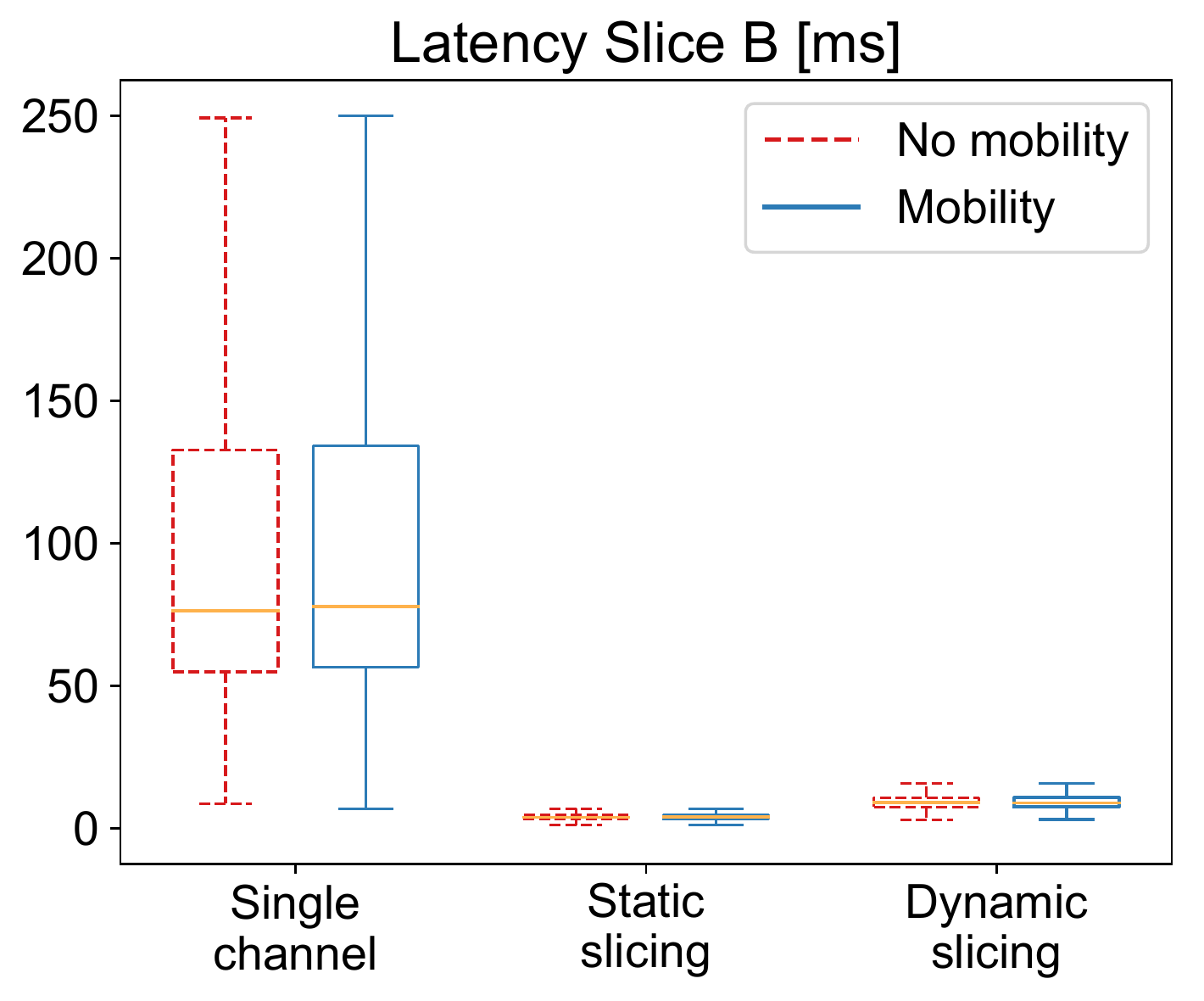}
    \caption{mMTC}\label{fig:latB}
    \end{subfigure}
    \begin{subfigure}[b]{0.32\textwidth}   
    \centering 
    \includegraphics[width=\columnwidth]{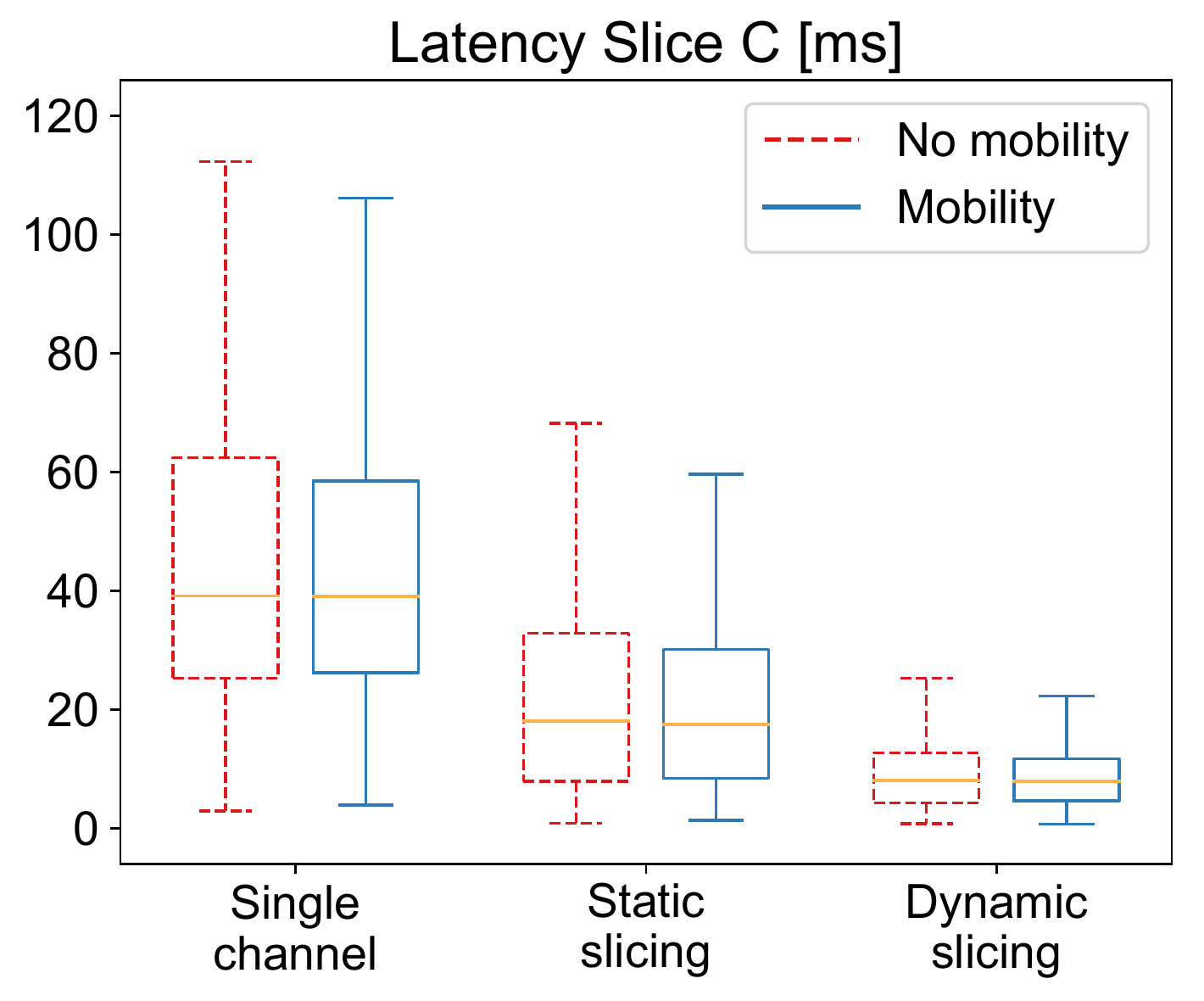}
    \caption{URLLC}\label{fig:latC}
    \end{subfigure}
\caption{Latency for the three slices A, B and C.
}
\label{fig:lat}
\end{figure*}



\subsection{Performance Analysis}

Following the \emph{ns-3} terminology a packet is considered transmitted when it is generated by the source, i.e. by the client application running on the \gls{sta}.
So, the value \emph{txPackets} for a traffic flow depends solely on the data rate for that flow.
For each flow, the packet error probability is calculated as:
\begin{equation}
P_{e} = \frac{txPackets - rxPackets}{txPackets}
\label{eqn:pe}
\end{equation}
where \emph{txPackets} and \emph{rxPackets} are the total number of transmitted and received packets within the flow, respectively.

Figure~\ref{fig:peA} reports the packet error probabilities experienced by \gls{embb} devices (Slice A).
The boxplots show the results obtained by merging data from the three introduced settings.
When network slicing is not implemented, and only one channel is utilised, the network becomes highly congested.
All the devices transmit on the same channel and send their \gls{rts} messages to the \gls{ap}.
The \gls{ap} can only serve one user at a time, and it will not reply with a \gls{cts} signal if it is busy.
The presence of many \gls{rts} messages creates interference which increases the error probability, that can be as high as 100\%.

Considering the results for Slice B (i.e.~\gls{mmtc} devices), plotted in Figure~\ref{fig:peB}, we see a significant improvement in the packet error probability.
With the single channel approach, the error probability is almost always lower than 0.2, even though long whiskers are present between the upper quartile and the $max$ value.
The packet error probability in Slice B is always zero when the static network slicing algorithm is considered.
On the other hand, considering the dynamic network slicing approach, we can see that the packet delivery performance is slightly worse than the single channel approach.
This is due to the adaptive transmission power scheme, which aims to save energy.
A lower transmission power results in a greater \gls{snr}, which worsens the error rate but allows for energy savings and that justifies the resulting performance deterioration.

Packet error probability experienced by \gls{urllc} devices is reported in Figure~\ref{fig:peC}.
When all the devices communicate through only one channel, we get the worst error probability, which reaches 100\%.
Conversely, network slicing techniques outperform by far the single channel approach.
The dynamic slicing algorithm achieves a slightly worse traffic delivery performance.
However, this algorithm employs a higher \gls{mcs} index to decrease the latency in the \gls{urllc} slice.

Let us now analyse the performance of the algorithms considering the \gls{e2e} latency experienced by the received packets in each slice.
This value is obtained from the \emph{delaySum} value (i.e.~the sum of all \gls{e2e} delays), for all the received packets of a flow, retrieved in \emph{ns-3} using the \emph{flowMonitor} module.
The average latency $L$ experienced by each packet in a flow is defined as: $L =delaySum/rxPackets$. 


Figure~\ref{fig:latA} shows the latency for \gls{embb} devices.
When all the devices communicate over the same channel, the experienced latency is poor.
It is difficult for the channel to accommodate the required total throughput and packets need to be queued.
In fact, they are sent only upon reception of the \gls{cts} signal from the \gls{ap}.
As a consequence, the average latency is raised up to almost \SI{2}{s} in some flows.
A deeper analysis of the results for static network slicing reveals that latency significantly decreases when the number of \gls{embb} devices is 2 and, and only slightly increases when there are 6 \glspl{sta} in this slice.
6 devices are likely to saturate the channel, leading to queued packets.
However, this problem is completely solved by the dynamic approach, which offers a flexible bandwidth scheduling.
By doubling the bandwidth, the channel capacity increases allowing a higher flow of packets with good performance.

The latency experienced in Slice B (i.e. by the \gls{mmtc} devices) is reported in Figure~\ref{fig:latB}.
Since the data volumes are lower, shorter queues are created and latency in this slice never reaches \SI{250}{ms}.
Again, slicing techniques clearly outperform the single channel method.
We notice that latency results for \gls{mmtc} users using the dynamic algorithm are slightly worse with respect to the ones obtained with static slicing.
Conversely, the objective of the dynamic algorithm for \gls{mmtc} is to use the lowest possible \gls{mcs} and consequently decrease the required transmission power.
In practice, in the performed simulations the dynamic slicing approach always used an \gls{mcs} strictly lower than 5.
The drawback of this is the increased latency, which for \gls{mmtc} devices is not crucial.

Finally, the last slice, characterised by \gls{urllc} services, is analysed in Figure~\ref{fig:latC}.
Overall, having a single channel which serves all the connected \glspl{sta} is the worst choice.
The dynamic algorithm clearly outperforms the other two thanks to the higher \gls{mcs} index used, without significantly compromising packet delivery performance.
In fact, it is designed to select the highest possible \gls{mcs} according to the bandwidth and the received powers.
This allows the transmition of more information in a shorter time.

In conclusion, the dynamic network slicing algorithm guarantees the lowest latency to \gls{embb} and \gls{urllc} devices, while maintaining a low packet error probability.
In addition, the dynamic slicing approach is also the preferred approach for \gls{mmtc} services since it is able to better manage latency and packet delivery in order to provide improved energy savings.

\subsection{Resource Utilisation Analysis}

The performance, both in terms of error probability and latency is greatly enhanced when slicing techniques are applied.
However, we want to investigate if this improvement is only possible because more resources are allocated, or if better efficiency can be achieved.
Thus, it is important to study the energy and bandwidth efficiency of our algorithms.

\begin{figure}
\centering
\includegraphics[width=0.67\columnwidth]{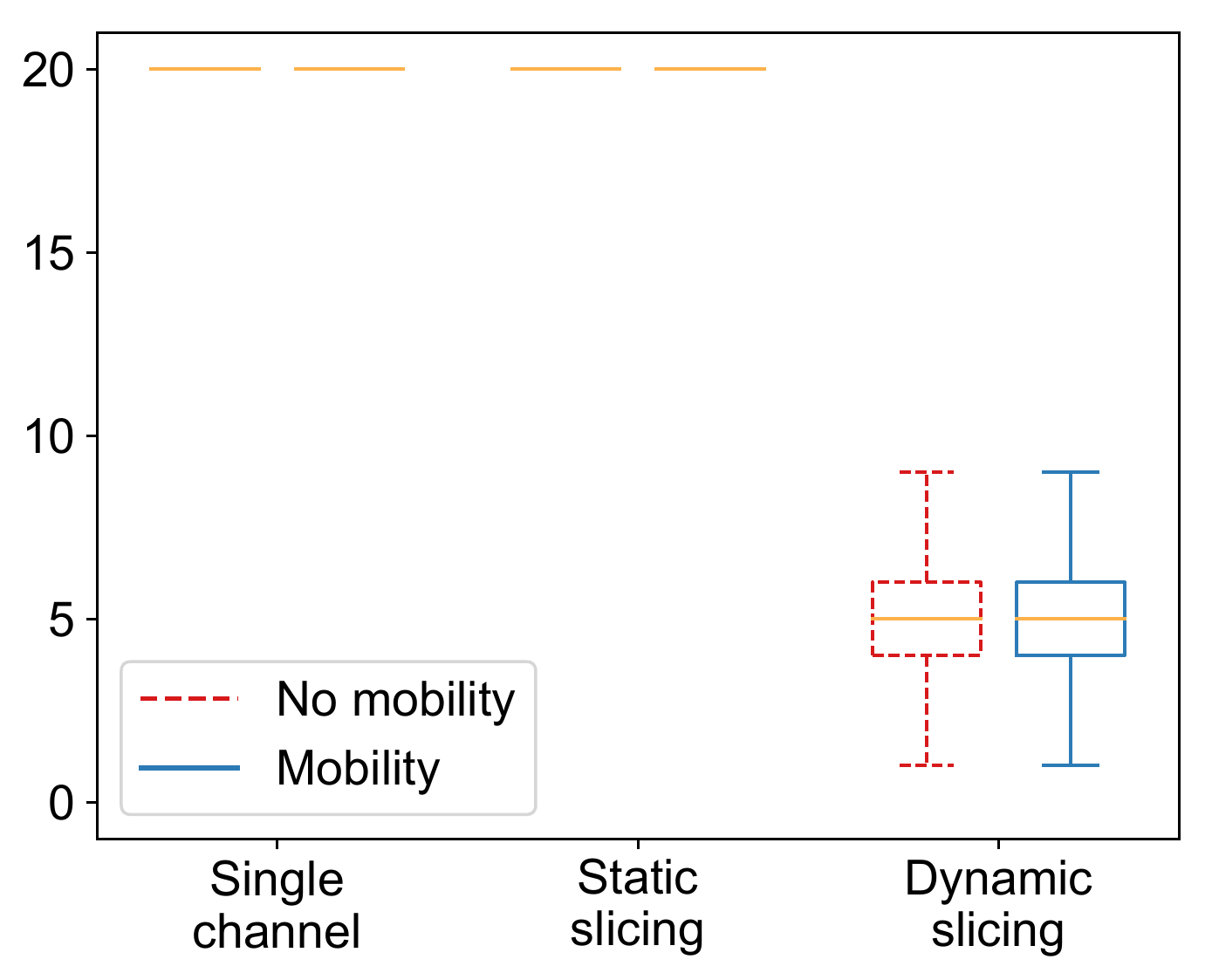}
\caption{Transmission power in Slice B (mMTC) [\si{dBm}].}
\label{fig:ptxB}
\end{figure}

For Slice B, which includes \gls{mmtc} devices, energy saving is the main concern.
For this reason, the dynamic network slicing algorithm aims at decreasing the required transmission power for these devices.
As shown in Figure~\ref{fig:ptxB}, the transmission power is lowered on average from \SI{20}{dBm} to \SI{5}{dBm} by using this algorithm.
In particular, considering the conversion from dB to linear units, a difference of \SI{15}{dB} means that the transmission power is decreased by $10^{15/10} \approx 32$ times in linear units.
Considering battery powered \gls{iot} devices served by the \gls{mmtc} slice, the dynamic network slicing approach could extend the battery life by up to 32 times.

Channel bandwidth is also a scarce resource that needs to be allocated by the algorithms, therefore we introduce the notion of spectrum efficiency.
This metric refers to the data rate that can be transmitted over a limited spectrum region.
First of all, we consider the total throughput achieved in each run, defined as the sum of the throughput of all the \glspl{sta}.
From the number of received packets on the $i$-th link, $rxPackets_i$, the total throughput can be obtained as:
\begin{equation}
Th_{sum} = \sum_{i=1}^{nSta} Th_i = \sum_{i=1}^{nSta} \frac{rxPackets_i \times pktSize \times 8}{simTime}
\end{equation}
where the parameters $pktSize$ and $simTime$ are given in Table~\ref{tab:wifin}.
$nSta=nStaA+nStaB+nStaC$ is the number of devices connected to the \gls{ap}, and $Th_i$ is the throughput of the $i$-th device.
After having defined the achieved throughput, we introduce the used bandwidth $B_w$ in each run.
This metric needs to be determined with three different expressions depending on which allocation algorithm is employed:
\begin{equation}
B_w = \begin{cases}
160, & \mbox{if single channel;} \\
B_{wA}+B_{wB}+B_{wC}, & \mbox{if static slicing;} \\
\frac{1}{N} \sum_{i=1}^{N} B_{wAi}+B_{wBi}+B_{wCi}, & \mbox{if dynamic slicing.} 
\end{cases}
\label{eqn:b}
\end{equation}

In Equation~\ref{eqn:b}, three cases are considered. One when a single channel is allocated with fixed bandwidth.
Another when the static network slicing algorithm is used, with three channel bandwidths defined at the beginning of the run $B_{wA}$, $B_{wB}$ and $B_{wC}$.
And finally, dynamic network slicing where in every interval $T$ new bandwidths are allocated.
Thus for each $i$-th interval with $i \in \{1, \ldots, N\}$ , where $N = simTime/T$, we average the assigned bandwidths $B_{wAi}$, $B_{wBi}$ and $B_{wCi}$.
Finally, the spectrum efficiency in each run is the ratio $\mu = Th_{sum} / B_w$:

\begin{figure}
\centering
\includegraphics[width=0.67\columnwidth]{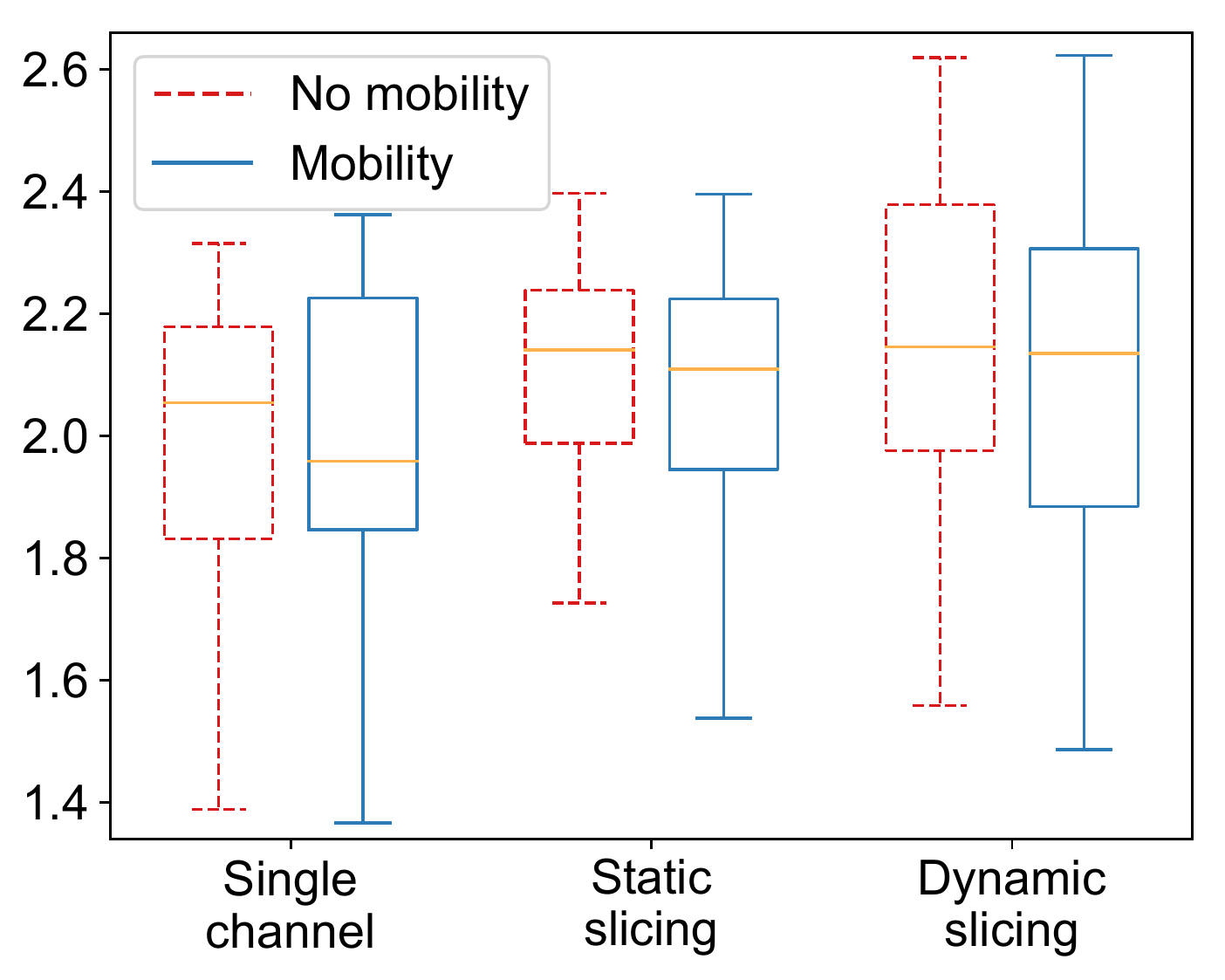}
\caption{Spectrum efficiency [\si{\bit\per\second/\hertz}].}
\label{fig:se}
\end{figure}

In Figure~\ref{fig:se}, the spectrum efficiency of the network is reported for each tested approach.
The data considered are no longer ``per flow'' as it was the case for the packet error probability and the latency.
Instead, we consider the data ``per run'', since the spectrum efficiency is calculated for each run.
Thus, the cardinality of the plotted datasets is lowered to 60 (i.e. three settings run with 20 seeds each), which results in longer whiskers in the box plots.

The two slicing approaches provide the highest spectrum efficiency.
Interestingly, even though the dynamic algorithm may periodically double the amount of used bandwidth, it has the best spectrum efficiency over the two other approaches, reaching up to \SI[per-mode=symbol]{2.6}{\bit\per\second/\hertz}.
Thus, we can state that allocating more bandwidth is not necessarily a waste of resources if carefully done.
Moreover, this confirms that resources can be adequately managed for different slices, taking into account not only their distinct requirements but also the overall performance of the system.

\section{Conclusion}
\label{sec:conclusion}

In this paper we show how network slicing can be implemented in Wi-Fi networks using different \glspl{ssid}.
We explore a slice setup based on the three main scenarios defined by 5G: \gls{embb}, \gls{mmtc} and \gls{urllc}.
We present two slicing algorithms where a network slice is allocated for each scenario, supporting multiple stations based on the expected performance regarding throughput, latency and energy efficiency.
One of the proposed slicing algorithms statically allocates three channels according to the expected throughput requirements, considering the theoretically necessary channel bandwidth.
The other, dynamically adapts the allocated channel resources at run-time, according to the network needs.

We validate the proposed approach and algorithms with extensive simulations using the \emph{ns-3} network simulator.
The obtained results reveal that our slicing approach outperforms today's access scheme in which an \gls{ap} serves all the connected users with a single wireless channel. 
We achieve lower packet error probabilities and lower latencies.
Furthermore, slicing is able to reduce the energy needed by low-power devices and increase the spectrum efficiency.
Thanks to these promising results, this study may pave the way for a future implementation of network slicing in Wi-Fi networks.

\bibliographystyle{IEEEtran}
\bibliography{main}

\end{document}